\documentclass[a4paper,11pt]{article}
\usepackage{jinstpub} 
\usepackage{parskip}
\usepackage{doi}
\usepackage[T5,T1]{fontenc}
\usepackage{svg}
\usepackage{comment}
\usepackage[backend=biber,style=numeric,sorting=none,doi=true,url=false,isbn=false,eprint=false]{biblatex}
\usepackage{xcolor} 
\usepackage{subcaption}

\hypersetup{
    colorlinks=true, 
    urlcolor=blue,   
    linkcolor=blue,  
    citecolor=blue   
}

\title{\boldmath Resolution Limiting Factors in Low-Energy Cascade Zenith Angle Reconstruction with the IceCube Upgrade}

\author[a]{Kaustav Dutta,}
\author[a]{Sebastian Böser,}%
\author[a]{Jan Weldert}
\author[b]{\textnormal{and} Martin Rongen}

\affiliation[a]{Institute of Physics, University of Mainz, Staudinger Weg 7, D-55099 Mainz, Germany}

\affiliation[b]{Erlangen Centre for Astroparticle Physics, Friedrich-Alexander-Universität Erlangen-Nürnberg, D-91058 Erlangen, Germany}

\emailAdd{kdutta@icecube.wisc.edu}

\abstract{The IceCube Neutrino Observatory includes low energy extensions such as the existing DeepCore subarray and the upcoming IceCube Upgrade, which will consist of seven new strings of photosensors with denser instrumentation than the existing array. The setup will allow for the study of neutrino oscillations with greater sensitivity compared to the existing instrumentation, improve neutrino mass ordering studies, and test for the unitarity of the PMNS mixing matrix with high precision. A critical component in these low-energy physics analyses is the accurate reconstruction of event information, particularly the zenith angle of incoming neutrinos. In this study, we discuss the processes that limit the zenith angle resolution, which include the transverse spread of the hadronic shower, in-ice photon scattering, module resolutions, and module noise. By considering approximations to these processes we aim to approach the intrinsic zenith angle resolution limits for neutral-current events characterized by a hadronic shower.}


\usepackage{ragged2e}
\usepackage[export]{adjustbox}

\begin{document}

\maketitle
\flushbottom
\setlength{\parindent}{20pt}

\section{Introduction}
\label{introduction}

The IceCube Neutrino Observatory consists of a cubic kilometer of ice instrumented with 5160 Digital Optical Modules (DOMs) distributed across 86 strings, positioned at depths ranging from 1450 m to 2450 m. Most of the detector features modules spaced 125 m apart horizontally and 17 m vertically, optimized for detecting astrophysical neutrinos with energies exceeding $\sim$100 GeV ~\cite{Aartsen:2013}. The bottom center of the detector includes 8 strings forming the DeepCore region where the optical transparency of the ice layers is the cleanest in the instrumented region ~\cite{Abbasi:2012}. Through its increased instrumentation density, DeepCore lowers the detection energy threshold to approximately 10 GeV ~\cite{Abbasi:2012} enabling improved event reconstruction of such low-energy events and allowing for the exploration of the fundamental properties of neutrinos ~\cite{Osc_analysis, numu_appearance, NSI_analysis, Sterile_analysis}.
\subsection{The IceCube Upgrade}
\indent The IceCube Upgrade ~\cite{ishihara2019icecube} will be deployed in the 2025/2026 Antarctic Field season, where seven additional strings of photosensors and calibration devices will be dispersed within the DeepCore volume. Most of the new optical sensors will be installed at depths between 2160 m and 2430 m, with a smaller number of calibration devices placed between 1300 m and 1900 m as well as below 2450 m. The modules will be positioned with an average horizontal and vertical spacing of 30 m and 3 m, respectively, designed to further lower the energy threshold to approximately 3 GeV ~\cite{eller2023sensitivity}. The increased instrumentation density will enhance photon statistics on a per-event basis and consequently improve event reconstruction at a few GeV. The majority of the new optical modules will be multi-PMT Digital Optical Modules (mDOMs) ~\cite{mDOM}, each equipped with 24 3" PMTs, along with several dual-PMT modules called DEggs\footnote{Dual optical sensors in an Ellipsoid Glass for Gen2} ~\cite{DEgg}. Compared to the currently deployed single-PMT modules and DEggs, the mDOMs offer enhanced directional coverage and nearly isotropic sensitivity allowing for more accurate reconstruction of neutrino directions.

\subsection{Event signatures}
\label{Event signatures}
\sloppy
\indent Neutrinos are detected by underwater or ice-based neutrino telescopes when they interact with the transparent detector medium or surrounding rock, producing relativistic charged particles. As these particles traverse the detector, they emit Cherenkov radiation which is recorded by the optical modules deployed within the detector~\cite{LU2020163678}. The events are classified as "cascades" or "tracks" depending on the light signature left in the detector by the produced charged particles. Above a few GeVs, neutrino-nucleon interactions are typically deep inelastic scatterings ~\cite{Bustamante:2019} which result in the disintegration of the nucleon and a subsequent hadronic shower, observable in all events starting inside the detector volume. The emitted Cherenkov photons undergo scattering as they propagate through the medium toward the modules which distorts their initial directional information. The finite number of PMTs in the receiving modules results in a finite photon directional resolution, which improves proportionally with the PMT count per module. Noise in the IceCube modules is non-correlated Poissonian distributed noise of a thermal origin and correlated noise originating from scintillation light due to natural radioactive decays within the module glass ~\cite{Larson_thesis}. These generate signals mimicking physics hits, further complicating the extraction of actual physics information.\\
\indent In $\nu_e$ CC and flavor-independent neutral-current (NC) interactions, electromagnetic and hadronic showers are created while $\nu_\mu$ charged-current (CC) interactions produce an additional muon which leaves a track signature. In $\nu_\tau$ CC interactions, a hadronic shower is formed at the interaction vertex, followed by either a hadronic or leptonic shower from the subsequent tau decay, with known branching ratios ~\cite{ParticleDataGroup:2020ssz}. At energies around 10 GeV, due to the coarse detector granularities, distinguishing between these different topologies becomes increasingly challenging ~\cite{osti_10349667}. In $\nu_e$ CC, $\nu_\tau$ CC and NC events, there are no experimentally observed differences due to negligible cascade elongation \cite{Steuer_theses} and tau decay \cite{nu_tau} lengths compared to the inter-string spacing. All of these events can thus be modeled well by a point-like single-cascade model in reconstructions which are described by relatively fewer parameters than muon tracks. Given the similarity of event topologies at these very low energies and motivated by the simpler event modeling, this study focuses on the directional reconstruction for NC events characterized by a hadronic shower.

\subsection{Importance of zenith angle reconstruction}
\label{imp_zenith}
\begin{figure}[!ht]
\centering
\includegraphics[width=0.6\textwidth]{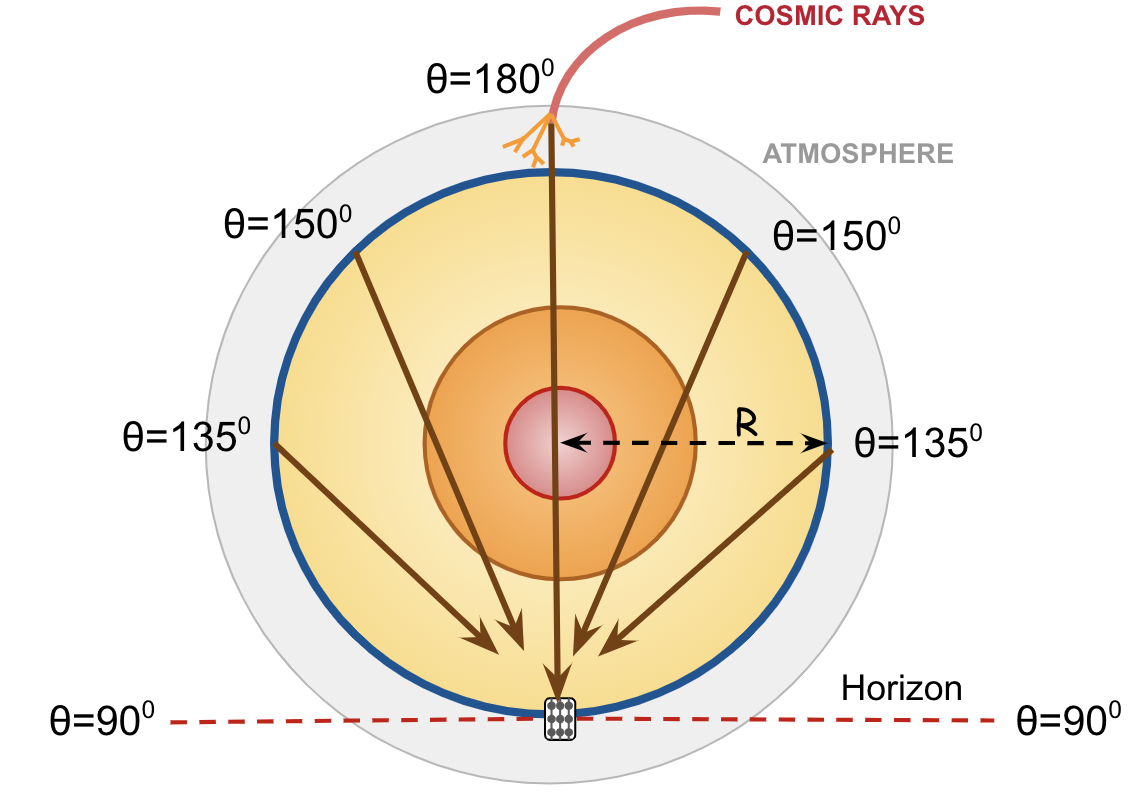}
\caption{Diagram illustrating the correlation between the distance a neutrino travels through the Earth and the incident zenith angle of the neutrino.}
\label{Zenith_L}
\end{figure}

\indent This study aims to analyze the processes that limit the zenith angle directional information in IceCube Upgrade events. This is directly relevant for all low-energy neutrino oscillation analyses in water (ice) Cherenkov experiments  ~\protect\cite{Osc_analysis} since it translates into the oscillation length $L$, given by $\textstyle L \approx 2R \lvert \cos \theta \rvert$ (where R denotes the radius of the Earth and $\theta$ is the zenith angle of the incoming neutrinos), as depicted in Fig.~\ref{Zenith_L}. Previous studies in IceCube have shown that a 5\%, 10\%, and 15\% improvement in the zenith angle resolution results in a 3\%, 10\%, and 15\% enhancement in Neutrino Mass Ordering (NMO) sensitivity, respectively ~\protect\cite{Jan_thesis}. Since better reconstruction yields improvements in analysis sensitivities, efforts are directed toward improving reconstruction algorithms. It is therefore essential to check the performance status of the reconstruction algorithms relative to the resolution limits, which are determined under the assumption of ideal detector modeling and by incorporating all physical processes that constrain the achievable resolutions. \\
\indent The study is motivated by a prior investigation pursued by the KM3NeT Collaboration\footnote{The Cubic Kilometre Neutrino Telescope (KM3NeT)} ~\cite{KM3NeT:2017},  which explored the information-limiting processes in $\nu_e$ CC and $\nu_{\mu}$ CC events in the ORCA\footnote{Oscillation Research with Cosmics in the Abyss (ORCA)} water Cherenkov detector. These included fluctuations in the hadronic shower composition, particle propagation, and photon sampling effects. The resolution limiting processes included in this study are the transverse shower spread, in-ice light scattering, module directional resolutions due to finite PMTs in the modules, and noise in the PMTs (Section \ref{results}). Given the lack of topological distinction between NC and CC cascade events at these low energies, the resolution limits derived in this study are equally applicable to $\nu_e$ CC and $\nu_\tau$ CC interactions, with only minimal corrections. The CC events show a strong zenith angle dependence because of their flavor specific nature and the baseline-dependent neutrino flavor oscillations. As a result, they contribute significantly to IceCube oscillation analyses \cite{Osc_2023, Osc_2018, Osc_analysis}.

\indent This paper is organized as follows. Section \ref{approximations} outlines the approximations used in the event simulation that make it more idealized in contrast to the Monte Carlo (MC) simulation chain used by the IceCube Collaboration. Section \ref{event_sim} details the event simulation procedure, including the incorporation of simplifications and information loss processes. In Section \ref{reco_method}, we describe the reconstruction methodology, the expected photon distributions in IceCube events, and the reconstruction resolutions based on the different measurements made by the optical modules. Section \ref{results} examines the impact of different processes that limit physics information in IceCube Upgrade events and evaluates the reconstruction performance using a toy simulation for a more accurate estimate of resolution limits. Finally, Section \ref{summary} summarizes the results and concludes the discussion.

\section{Approximations in this study}
\label{approximations}
\indent Since events in Cherenkov detectors are characterized by detected photon signatures, the reconstruction of events relies on the precise modeling of the medium through which the photons propagate. In the IceCube experiment, the optical properties of glacial ice show a strong depth dependence which reflects the changes in Earth's climate history ~\cite{Aartsen:2013}. The ice sheet is also not horizontally uniform and is titled due to the underlying bedrock topography ~\cite{chirkin2023improved}. In addition, there is a direction-dependent anisotropy effect due to birefringent polycrystals which leads to an average deflection in the diffusion pattern towards the flow axis ~\cite{IceCube_ice}. Moreover, the modules are deployed into holes made by hot water drilling. This hole ice ~\cite{fiedlschuster2019effect}, on re-freezing, has different optical properties than the bulk ice and contains a bubbly column (with poor optical properties) approximately in the center of the column.\\
\indent Due to these ice anisotropies and the grid-like geometry of the detector, the measured angular and timing distribution of photons (Fig.~\ref{PDF_disagreement}) by a receiving optical module depends strongly on the interaction vertex position and event direction. The similarity of the scattering length and module spacing scales further make the photon distributions analytically intractable. In an ideal likelihood-based reconstruction, the actual expected
photon distributions must be generated for each optical module for every hypothesized position
and direction of the event. Given the detector’s extensive volume and the large number of optical
modules, this results in a substantial simulation campaign and makes the ansatz computationally
infeasible. In practical application, simplified photon distribution models using approximate symmetries and translational invariance are employed \cite{improvedcascadereco:2024csv}. In this study, in contrast, we circumvent the modeling challenges by simplifying the assumptions about the photon propagation and using other approximations in the simulation of the detector.\\
\indent Events are simulated and reconstructed in a homogeneous ice medium, where the optical properties remain uniform throughout the volume, incorporating translational and rotational symmetry in the photon distributions in space and time. There is no ice stratification, layer undulations, birefringence, or hole ice, which otherwise require detailed parameterization of the ice-depth dependence. All PMTs in the modules are assumed to be identical, with the same angular acceptance, which defines the fraction of photons detected at a given angle relative to the PMT axis. Moreover, the simulations in this study only include 24-PMT mDOMs at all module positions in the detector geometry whereas typical MC simulations also include other module types such as pDOMs and DEggs. This results in better directional resolutions since mDOMs offer a better pointing resolution as compared with other modules with fewer PMTs. The assumption is appropriate for this study which seeks to estimate optimistic resolution limits. \\
\indent The chosen detector geometry is significantly larger than the actual IceCube Upgrade volume, with the events completely confined within the detector which prevents emitted photons from escaping and loss in directional information. In the actual detector, however, events at the edge of the detector emit photons that are undetected which may degrade the reconstructed resolution. The strings are arranged in a hexagonal geometry where the vertical and horizontal module separations are fixed at 30m and 3m, respectively. In contrast, the actual Upgrade array will have non-uniform string spacings ranging between 20m and 40m and will include a DeepCore and IC-79 string with sparser instrumentation. Because Upgrade strings have denser instrumentation and more PMTs in each mDOM, they register significantly more photon hits than non-Upgrade strings. As a result, the influence of photon hits on non-Upgrade strings on the final resolutions is negligible.\\
\indent To generate the expected angular and timing distribution of detected photons relative to the shower axis and vertex time (Section \ref{para_estimation}), respectively, a randomized geometry (Section \ref{photon_dist}) is created. Events are simulated with interaction vertices distributed uniformly throughout this geometry. The vertex-averaged distribution of detected photons across all modules removes the photon sampling bias caused by the local geometry around the event. This approach allows the use of a single probability density function (PDF), representing the modeled angular and timing distribution of photons summed over all hit modules and averaged over multiple events, to reconstruct all events. The PDFs generated in this way are independent of the source-receiver distance and represent an idealization, deviating from standard likelihood reconstructions (Section \ref{event_reco}) that rely on photon distributions from each module. The expected number of photon hits in a module, defined as charge in this study, is distance-dependent and has been incorporated in the likelihood formulation. Note that the definition of charge in this study differs slightly from the one used by the IceCube Collaboration ~\cite{SPE_template}, as explained later in this section.\\
\indent Although detailed modeling of angular and timing profiles of hit photons from each sensor adds event information which may lead to more accurate reconstructions, they demand large lookup tables generated with high statistics MC simulations that consume substantial memory. The azimuthal symmetry around single PMT modules, which previously reduced the dimensionality of photon tables by one dimension, is disrupted by the segmented mDOMs. This symmetry breaking increases the size of these photon tables to hundreds of gigabytes, leading to substantial data management challenges that this study seeks to avoid. A rough estimate of the potential resolution improvements by including the per-module photon information is presented in Section \ref{results}.\\
\indent The MC simulation chain (Section \ref{results}) used by IceCube is further idealized by omitting the conversion of incident photons into PMT readout pulses. In the official simulation, the pulse charge is sampled from a single photoelectron (SPE) charge spectrum generated at a standard PMT gain of 10$^7$ for an in-ice module ~\cite{SPE_template}. If the sampled charge exceeds 0.25 PE, discriminators trigger the data acquisition; otherwise, the photon is not registered resulting in a loss of event information. This study avoids such losses by directly using the incident photon data, bypassing PMT and electronic effects. Additionally, the idealized simulation excludes the complexities of event filtering and advanced pulse-cleaning algorithms based on photon hit distributions. In the standard simulation chain, pure noise events are filtered out in the event selection stage by applying a cut on the surviving pulse count after noise-cleaning. These steps are taken to ensure that a high-purity sample of neutrino events, consisting mostly of physics hits, is reconstructed. This study replicates the process by first simulating physics hits from a neutrino event and then introducing noise hits based on the published dark rate of mDOMs ~\cite{mDOM_dark_rate}, achieving the same signal purity as the standard simulation, as detailed in the next section.

\section{Event Simulation}
\label{event_sim}
\begin{figure}[!ht]
\centering
\includegraphics[width=0.6\textwidth]{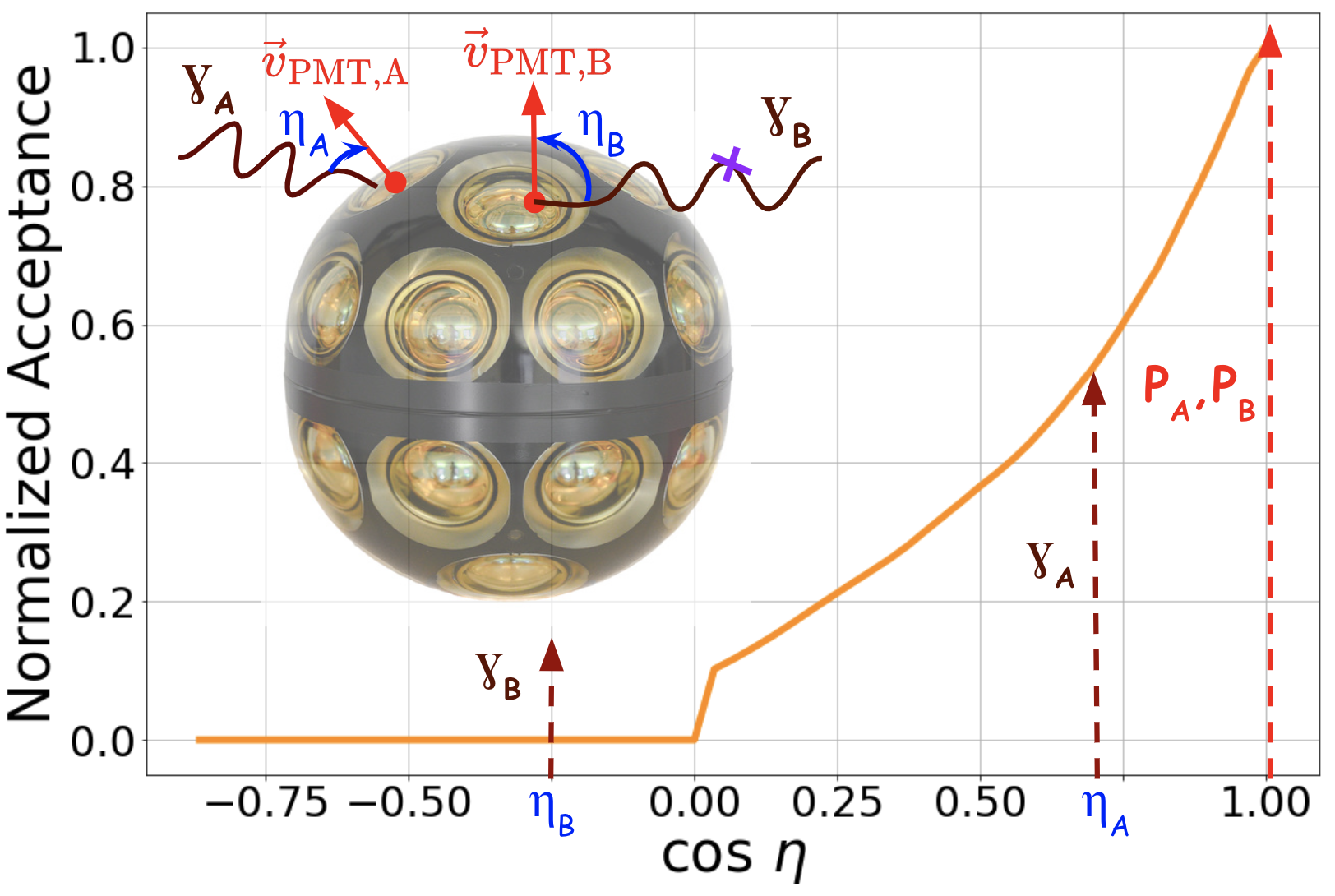}
\caption{Schematic illustrating the incorporation of module resolutions in the simulation. $\gamma_\text{A}$ and $\gamma_\text{B}$ are the MC truth direction of two photons, and the angles between these photons and corresponding nearest PMT axis vectors, respectively, are indicated by P$_\text{A}$ and P$_\text{B}$ respectively. The purple cross indicates that $\gamma_B$ does not register a photon hit due to the low detection efficiency of the PMT at larger angles.}
\label{mDOM_simulation_2}
\end{figure}

This study uses the standalone, open-source version of the Photon Propagation Code (PPC) ~\cite{PPC} used by the IceCube Collaboration. Events are simulated as purely hadronic cascades resulting from NC interactions mediated by a neutral boson, with no charged lepton emitted at the interaction vertex. The standard input to PPC is comprised of information on an IceCube event which includes the event type, interaction vertex coordinates and time, incoming directions in zenith and azimuth, and the energy of the hadronic shower. Since only the directions of outgoing charged particles within the detector are experimentally observable, the study focuses exclusively on the intrinsic directional resolution limits. The uncertainties associated with the kinematic opening angle between the incoming neutrino and shower axis are not included. The average kinematic angle between the incoming neutrino and the shower axis varies with energy ~\cite{kinematic_angle:2022} approximately as $\Psi_{\hat{\nu},\vec{u}} \approx \alpha \left( \frac{E_{\nu}}{\text{GeV}} \right)^{-\beta}$. The coefficients $\alpha$ = 12$^\circ$ and $\beta$ = 0.3 were fitted to the angular distribution of shower secondaries in 10 GeV purely hadronic cascades using the GENIE ~\cite{Andreopoulos:2010} event generator.\\
\indent The charged particles in the shower are generated and propagated using GEANT3 simulation software \cite{Geant3} (version 3.16). The hadronic interactions within ice are modeled with the GHEISHA 7.0 package \cite{GHEISHA} integrated within the GEANT3 framework. The Cherenkov photon yield is determined based on the energy loss of the charged particles, and the parameterisations ~\cite{Wiebusch:1995bw} are directly used by PPC to determine the distribution of Cherenkov photons. The photons are then propagated by PPC through the ice until they get absorbed or reach one of the modules. The scattering angle of propagating photons at each scattering point is sampled from a linear combination of scattering functions ~\cite{Aartsen:2013}, which is predominantly characterized by forward scattering. The output from PPC includes the strings and modules hit by photons with the corresponding timestamps, photon directions at the point of incidence on a module, and the polar coordinates of the impact point relative to the DOM center.\\
\indent The overall angular distribution of photons emitted by the individual charged particles in the hadronic shower relative to the shower axis, referred to as "shower spread" in this study, is given as $\frac{dl}{dx} \approx \exp(-{2.61} \cdot x^{0.39}) \cdot x^{-{0.61}}$, where $l$ represents the length of the light-emitting track segment. The variable $x$ is defined as $x$ = 1-cos($\delta$) ~\cite{Wiebusch:1995bw}, where $\delta$ denotes the angle between the emitting charged particle track and emitted photon direction. To switch off the shower spread, the simulation skips the step in the simulation that samples the emission angles of photons from the above-described distribution and instead draws them from a Cherenkov cone, whose axis aligns with the shower axis, with a wavelength-dependent cone angle ~\cite{PhysRev.52.378}. The scattering and absorption of photons are governed by the bulk ice properties and have been derived from ~\cite{Aartsen:2013}. For this study, the scattering and absorption coefficients are set to $a_e = 0.0013 \, \mathrm{m}^{-1}$ and $b_e = 0.013 \, \mathrm{m}^{-1}$, respectively. These values represent the ice properties at a depth of approximately 2300 m, corresponding to the most transparent ice layers within the Upgrade physics region.\\
\indent The PPC simulations only involve the generation and propagation of photons through the ice. The PPC output is adjusted to include other information loss factors such as module resolutions and module noise. To simulate a multi-PMT Digital Optical Module (mDOM) ~\cite{mDOM}, which will be the most common optical module in the Upgrade, 24 PMTs are arranged in four belts positioned at zenith angles of $\pi/8$, $3\pi/8$, $5\pi/8$, and $7\pi/8$ relative to the center of the module. The PMT population per belt is 5, 7, 7, and 5, respectively, with the PMTs within each belt uniformly distributed in azimuth. The overall angular resolution of the optical modules, referred to as "module resolution" in this study, is simulated using the following procedure to account for the finite number of PMTs and the shape of the angular acceptance curve for each PMT. \\
\indent The PMT closest to the point of photon impact is first chosen as shown in Fig.~\ref{mDOM_simulation_2}. The angle between the incident photon directions and the PMT axis vectors, such as $\eta_\text{A}$ or $\eta_\text{B}$, is then calculated. Rejection sampling is employed to select photons based on this angle and the angular acceptance curve of a PMT. For our reconstruction, PMT axis vectors, such as  $\vec{v}_\text{PMT, A}$ or $\vec{v}_\text{PMT, B}$, are assumed as the photon directions for the sampled photons reflecting the known polar orientation of the hit PMTs. The procedure is illustrated in Fig.~\ref{mDOM_simulation_2} using two photons, \(\mathrm{\gamma_A}\) and \(\mathrm{\gamma_B}\), incident on a multi-PMT module. The angle \(\mathrm{\eta_A}\) between \(\mathrm{\gamma_A}\) and the nearest PMT vector $\vec{v}_\text{PMT, A}$ is smaller than the angle \(\mathrm{\eta_B}\) between \(\mathrm{\gamma_B}\) and $\vec{v}_\text{PMT, B}$. Consequently, \(\mathrm{\gamma_A}\) has a higher probability of registering a photon hit as the acceptance function (shown in orange) increases monotonically with decreasing angles. The simulation method here does not preserve the absolute detection efficiency ~\cite{Abbasi:2010} of the mDOMs. However, this detail is inconsequential for our study because all resolutions are shown as a function of photon hit counts. Whenever module resolutions are not included in the simulation, the true direction of the photon is used instead.\\

\indent The signal purity of an IceCube event is defined as the fraction of the total observed photon hits that are physics hits, after noise cleaning. The total observed photon hits also include noise hits which are classified as correlated noise hits and uncorrelated noise hits. The overall dark rate of each PMT in mDOMs was measured to be 750Hz ~\cite{mDOM_dark_rate} which translates to 0.1 noise hits per module in a waveform readout window of approximately 6$\mu$s. As a first-order approximation, the number of noise hits for a module is assumed to be Poisson distributed with $\mu$=0.1. This corresponds to a high signal purity of approximately 90\% which is in agreement with the values given in ~\cite{eller2023sensitivity}. To simulate a noise hit in a module, a PMT is randomly selected and a timestamp is sampled uniformly within the event time window. The overall noise rate is included in the simulation without distinguishing between correlated and uncorrelated noise and therefore, without any spatial or temporal clustering of noise hits. The generated noise hits are then combined with the physics hits to give the total observed hits for an event. Given the negligible impact of noise on our reconstruction (Section \ref{results}) due to the high signal purity, this technique provides a sufficiently good approximation of noise contamination in IceCube events.

\section{Reconstruction Methodology}
\label{reco_method}

\subsection{Reconstruction techniques in IceCube}
\label{event_reco}
The longitudinal length of hadronic showers in water and ice increases logarithmically with the shower energy and measures 2-4 meters at energies of O(10 GeV) ~\cite{Steuer_theses, R_del:2013, shower_KM3NeT}. Since the sparse instrumentation of neutrino telescopes does not allow the resolution of shower details at such low energies, they are modeled as point-like sources with an anisotropic photon emission. A point-like cascade event in IceCube is characterized by 6 parameters ~\cite{Abbasi:2022}: the incoming zenith angle $\theta$ and azimuth $\phi$ of the shower axis, interaction vertex coordinates $\vec{v}$ = \{v$_x$,v$_y$,v$_z$\} and the vertex time t$_v$. The average number of emitted photons scales linearly with the summed track length of all charged particles, which increases linearly with the energy of the particle ~\cite{RADEL2013102, Wiebusch:1995bw}. This results in a sparse pulse distribution, with very few detected photons per event at energies of a few GeV ~\cite{Abbasi:2022}. Noise cleaning further reduces the physics hits due to misclassification posing challenges to the reconstruction process and necessitating an accurate algorithm.\\
\indent An IceCube reconstruction algorithm employed in the low-energy regime is the \textit{Single-string ANTares-inspired Analysis} (SANTA) ~\cite{Abbasi:2022} which utilizes information from minimally scattered photons. This is done by a rigorous photon hit selection based on the delay timing of photons followed by an event selection based on the surviving photon hit counts. A chi-squared fit on the measured arrival photon hit times for an event hypothesis is finally performed to obtain the optimized parameters. Another standard reconstruction, which additionally exploits the in-ice scattering information from all detected photons, is a likelihood algorithm ~\cite{Abbasi:2022} known as \textit{Reverse Table Reconstruction} (RETRO). This method fits a point-like cascade model to the measured light yield in the detector, where the cascade template outlines the expected light distribution in a module for the given hypothesis parameters. A binned Poisson likelihood is then used to determine the hypothesis that best describes the observed light pattern. The likelihood-based approach provides superior resolution in all reconstructed directions but comes at a significantly higher computational cost due to reasons discussed in Section \ref{approximations}.

\subsection{Reconstruction framework for this study}
\label{para_estimation}
From the discussions in Section \ref{Event signatures} and \ref{event_reco}, we model the hadronic cascades as point-like sources in this study. The zenith angle $\theta$, a key parameter for oscillation analyses, is determined by reconstructing the direction unit vectors $\vec{u} = \{ u_x, u_y, u_z \}$ of the shower axis. Combined with the parameters listed in the previous section, the study aims to reconstruct 7 parameters - $u_x, u_y, u_z, v_x, v_y, v_z, t_v$ - sufficient to describe the shower topology. \\
\indent The reconstruction uses an extended unbinned likelihood approach for its statistical robustness, with the assumption that it achieves minimum variance bounds since the assumed PDFs accurately represent the true distributions. Our method uses the direction, residual timing, and charge information of the detected photons, followed by a combined likelihood that exploits the correlation between these observables which is finally used for reconstructions in Section \ref{results}. For an anisotropic point source, the angular distribution of the incident photons $\vec{\gamma}$ relative to the direction of the shower axis $\vec{u}$ depends on the shower axis direction because more photons are detected in the modules in front of the shower. For the same reason, the number of photons registered by sensors is direction-dependent because sensors in front of a shower receive more photons than those behind it. The residual timings, defined as the difference between the observed $t_\text{obs}$ and geometric $t_\text{geo}$ timestamps, are sensitive to shower direction since sensors behind the shower receive more backscattered and delayed photons. \\
\indent In the likelihood reconstruction, events are first simulated by including a specific information loss process. The expected PDFs within the likelihood description are then generated using the simulated events. As a result, these distributions vary depending on the processes included in the simulation. For instance, a simulation that accounts for module resolutions projects the true directions of incident photons onto the nearest PMT axis, resulting in more smeared angular distributions of hit photons compared to those using the true incident directions. These events are next individually reconstructed using the generated PDFs. To reconstruct a single event, the likelihood functions used in this study are described as follows:
\begin{equation}
L_d(\boldsymbol{\theta} \mid \Delta \Psi, \Delta \zeta) = \prod_{s=1}^{N_{\text{sens}}} \prod_{i=1}^{N_s} p_d(\Delta \Psi_{i,s}, \Delta \zeta_{s} \mid \boldsymbol{\theta})
\quad\text{(photon direction information)},
\label{LLH_dir}
\end{equation}
\begin{equation}
L_t(\boldsymbol{\theta} \mid \Delta t, \Delta \zeta) = \prod_{s=1}^{N_{\text{sens}}} \prod_{i=1}^{N_s} p_t(\Delta t_{i,s}, \Delta \zeta_{s} \mid \boldsymbol{\theta})
\quad\text{(photon timing information)},
\label{LLH_time}
\end{equation}
\begin{equation}
L_q(\boldsymbol{\theta} \mid {N_s}) = \prod_{s=1}^{N_{\text{sens}}} p_q({N_s} \mid \boldsymbol{\theta})
\quad\text{(charge information)},
\label{LLH_charge}
\end{equation}
\begin{equation}
L_\text{joint}(\boldsymbol{\theta} \mid \Delta \Psi, \Delta t, \Delta \zeta, {N_s}) = \prod_{s=1}^{N_{\text{sens}}} \prod_{i=1}^{N_s} p_{\text{joint}}(\Delta \Psi_{i,s}, \Delta t_{i,s}, \Delta \zeta_{s} \mid \boldsymbol{\theta}) \cdot p_q({N_s} \mid \boldsymbol{\theta})
\quad \text{(correlated information)},
\label{LLH_combined_corr}
\end{equation}

where $\boldsymbol{\theta} = \{\vec{u}, \vec{v}, t_v\}$ is the 7-dimensional parameter set of the event, $s$ is the module index, $i$ is the photon index for a given module index $s$, and N$_\text{sens}$ denotes the total number of modules. The angle enclosed between the shower axis $\vec{u}$ and a detected photon $\vec{\gamma}_{i,s}$ is represented by $\Delta \Psi_{i,s}$, and $\Delta \zeta_{s}$ represents the angle between the shower axis and a vector $\vec{r}$ connecting the vertex and a module. The residual timing of a detected photon, defined as the difference between observed and geometric timestamps, is denoted by $\Delta t_{i,s} = t_{\text{obs},i,s} - t_{\text{geo},i,s} (\vec{\theta})$. The photon direction, timing, and their joint PDFs, $p_d$, $p_t$, and $p_\text{joint}$ are the expected averaged distributions over all hit sensors and are not individually defined for each sensor. The photon count per module, or charge, follows a Poissonian distribution given by
\[
p_q(N_s \mid \boldsymbol{\theta}) = \frac{\mu(\boldsymbol{\theta})^{N_s}}{N_s!} e^{-\mathcal{N}(\boldsymbol{\theta})},
\]
where $\mu$ and $N_s$ represent the expected and observed charge for a specific module. The normalized expected charge for the hit modules in an event is shown in Fig.~\ref{PDF_charge}. The generation procedure of these PDFs is further discussed in Section \ref{photon_dist}. 

\begin{figure}[!ht]
\centering
\begin{subfigure}{0.49\textwidth}
    \includegraphics[width=\textwidth]{Images/direction_info_schematic.pdf}
    \caption{}
    \label{fig:direction_info}
\end{subfigure}
\hfill
\begin{subfigure}{0.49\textwidth}
    \includegraphics[width=\textwidth]{Images/timing_info_schematic.pdf}
    \caption{}
    \label{fig:timing_info}
\end{subfigure}
\caption{Illustrative diagram showing the (a) directional information and (b) timing information of the shower axis direction contained in detected photons. Vertex coordinates and vertex time are denoted by $\vec{v} = \{v_x, v_y, v_z\}$ and $t_v$, respectively, while all photon direction vectors are shown as $\vec{\gamma}$. The green arrow shows the axis vector $\vec{v}_\text{PMT}$ of the nearest PMT from the incident photon. The shower axis is indicated by $\vec{u}$=$\{u_x,u_y,u_z\}$. The angle between the detected photons and the shower axis is $\Delta\Psi$ while the angle between the radius vector $r$ and the shower axis is $\Delta\zeta$. The angle between the PMT vector $\vec{v}_\text{PMT}$ and incident photons is indicated by $\Delta\eta$. The Cherenkov angle is denoted by $\theta_\text{c}$.}
\label{fig:angular_distr_bare_scatt}
\end{figure}

\indent Equation \ref{LLH_dir} calculates the likelihood of observing a given angular distribution $\Delta \Psi$ of detected photons $\vec{\gamma}$ and hit modules relative to the shower axis $\vec{u}$ (see Fig.~\ref{fig:direction_info}) for different parameter hypotheses $\boldsymbol{\theta}$. This formulation exploits the radial symmetry of the photon emission relative to the shower axis $\vec{u}$ to infer its direction. When module resolutions are considered, the photon direction is replaced by the axis direction $\vec{v}_\text{PMT}$ of the hit PMT. Equation \ref{LLH_time} evaluates the likelihood based on the observed residual timing $\Delta t$ of detected photons along with the angular distribution $\Delta \zeta$ of hit modules (see Fig. \ref{fig:timing_info}). Equation \ref{LLH_charge} determines the likelihood of observing a charge in a module at a radial distance $r_s$ from the vertex $\vec{v}$ and an angle $\Delta \zeta$ relative to the shower axis $\vec{u}$, based on an expected charge template (Fig.~\ref{PDF_charge}). Finally, the combined likelihood is calculated by simultaneously taking into account the observed photon directions, residual timing, angular distribution of hit modules, and charge in a module (Eq.~\ref{LLH_combined_corr}). The likelihood product is calculated over per-module detected photons N$_s$ (except Eq.~\ref{LLH_charge}) and then over hit modules N$_\text{sens}$, and the hypotheses that minimize the negative log-likelihood $\mathcal{L}$ are identified as the reconstructed parameters. The minimization is achieved with the Nelder-Mead method ~\cite{Nelder_Mead} with the truth parameter values as the initial seed in order to avoid bias from incomplete minimization. The parameters are illustrated schematically in Fig.~\ref{fig:angular_distr_bare_scatt} where only the scattering effects are shown on the directionality and temporal distributions of observed photons relative to unscattered photons. 

\subsection{Photon distributions}
\label{photon_dist}

\indent The likelihood reconstruction described in Section \ref{para_estimation} is dependent on the accurate modeling of the underlying PDFs. Due to the grid-like geometry of the detector, the direction, timing, and joint PDFs, $p_d$, $p_t$, $p_\text{joint}$, and the expected module charge $p_q$ in a homogeneous ice medium are dependent on the interaction vertex position and direction hypothesis of the shower axis. This is shown in the left panels of Fig.~\ref{PDF_disagreement} for two combinations of incoming zenith angle $\theta$ and azimuth $\phi$ of the shower axis for a fixed vertex position and the corresponding angular (top) and temporal (bottom) distributions of detected photons relative to the shower axis. For this study, we shall refer to them as the true integrated PDFs that model the integrated, or cumulative 
angular and timing distribution of photons over all hit modules when each event parameter is fixed to the true value. 

\begin{figure}[!ht]
\centering
\includegraphics[width=1.0\textwidth]{Images/PDF_disagreement.pdf}
\caption{\textbf{Left panels:} Contours of two-dimensional KDE fits (solid lines) to the angular (top) and timing (bottom) distributions of hit photons for a single event with a fixed vertex position, comparing two different incoming direction hypotheses; \textbf{Middle panels:} KDE fits to the angular (top) and timing (bottom) distributions of hit photons for the same hypotheses but averaged over multiple vertex positions; \textbf{Right panels:} Pull distribution showing the agreement between the 2D histograms for the two direction scenarios as shown in the middle panel.}

\label{PDF_disagreement}
\end{figure}

\begin{figure}[!ht]
\centering
\begin{minipage}{0.35\textwidth}
    \centering
    \includegraphics[width=\textwidth]{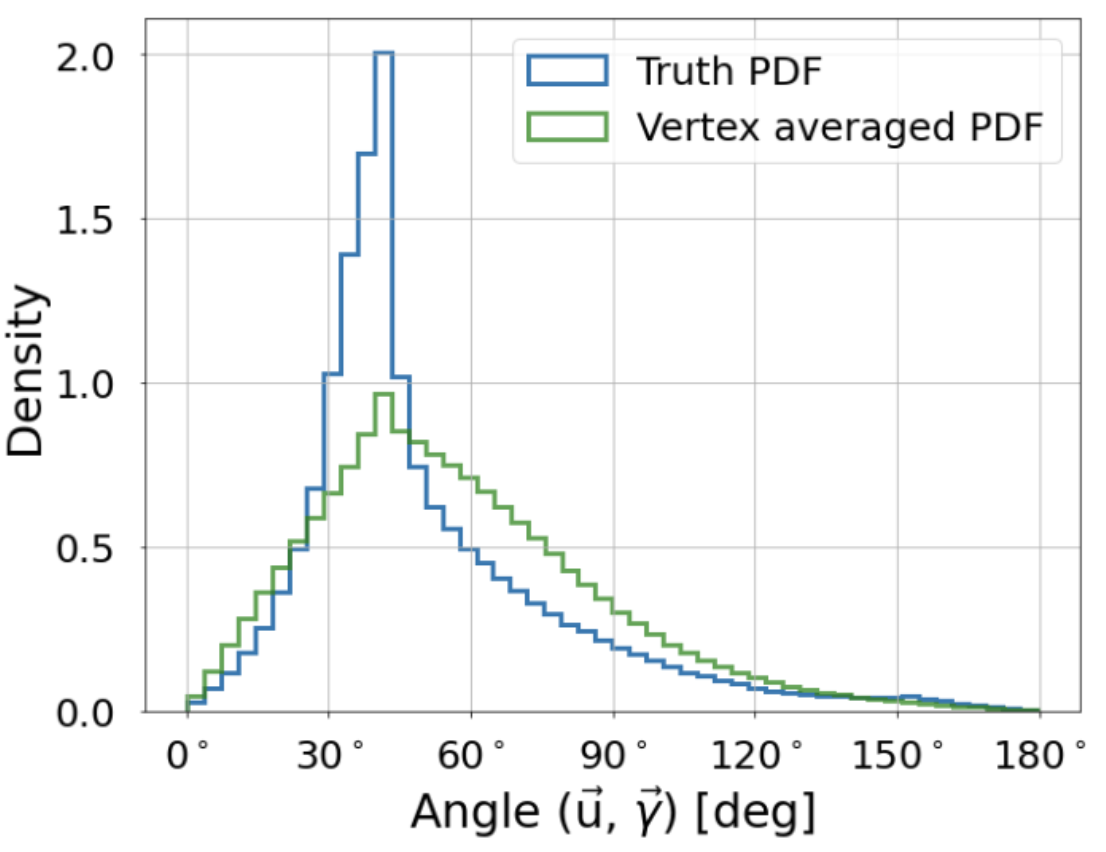}
    \subcaption{}
\end{minipage}
\begin{minipage}{0.36\textwidth}
    \centering
    \includegraphics[width=\textwidth]{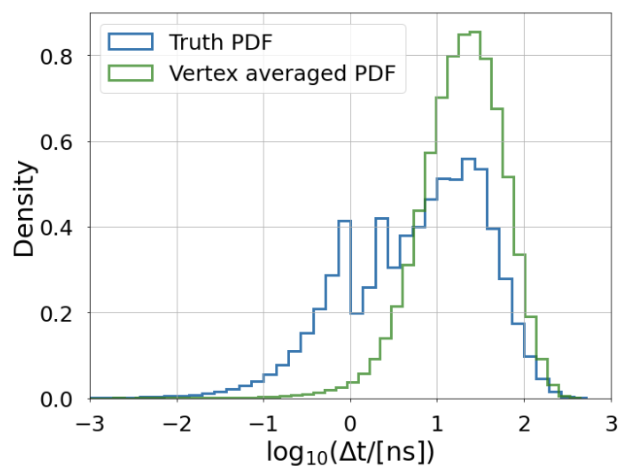}
    \subcaption{}
\end{minipage}
\begin{minipage}{0.27\textwidth}
    \centering
    \includegraphics[width=\textwidth]{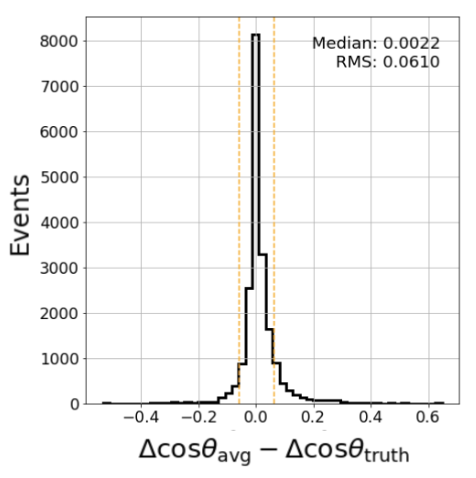}
    \subcaption{}
\end{minipage}
\caption{Reconstruction performance with truth (blue) and average (green) integrated PDFs when a vertex is 1 m away from the closest module and the event direction is towards the module in an Upgrade-like module spacing; (a) Angular distribution of detected photons relative to the shower axis, (b) Residual timing distributions of detected photons, and (c) Distribution of differences in zenith angle reconstruction resolutions, $\Delta\cos\theta_\text{avg}$ and $\Delta\cos\theta_\text{truth}$, when events are reconstructed with the averaged and truth PDF, respectively.}
\label{case_close}
\end{figure}

\begin{figure}[!ht]
\centering
\begin{minipage}{0.35\textwidth}
    \centering
    \includegraphics[width=\textwidth]{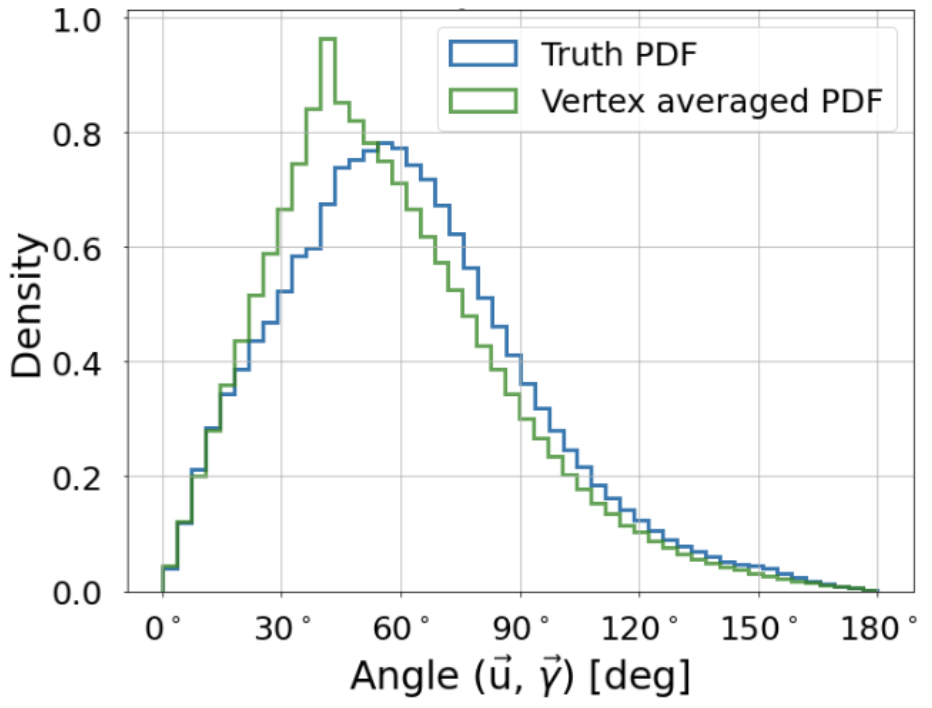}
    \subcaption{}
\end{minipage}
\begin{minipage}{0.36\textwidth}
    \centering
    \includegraphics[width=\textwidth]{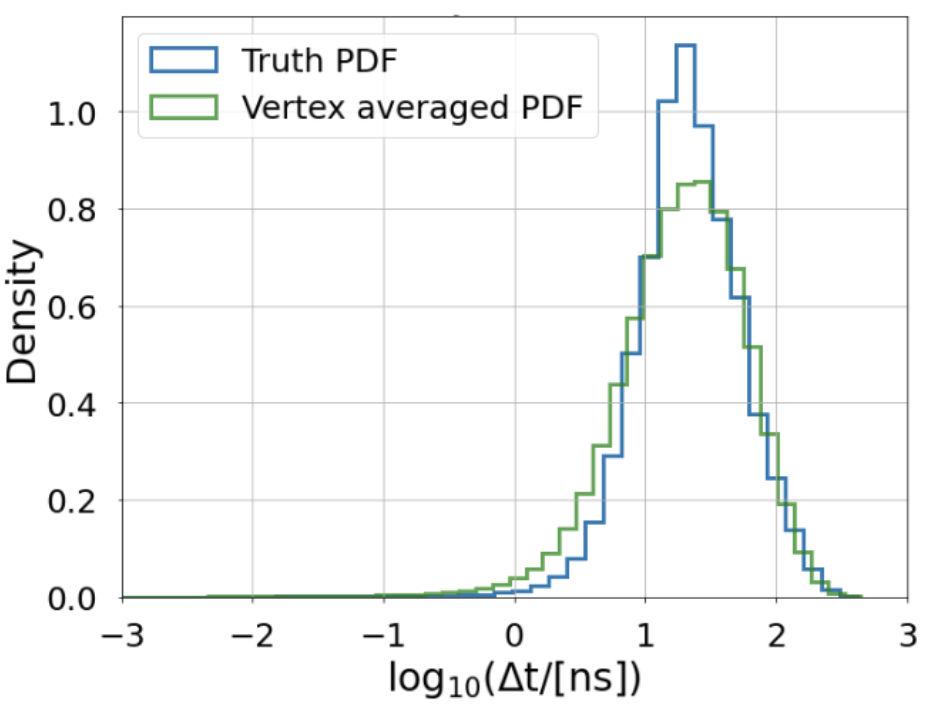}
    \subcaption{}
\end{minipage}
\begin{minipage}{0.265\textwidth}
    \centering
    \includegraphics[width=\textwidth]{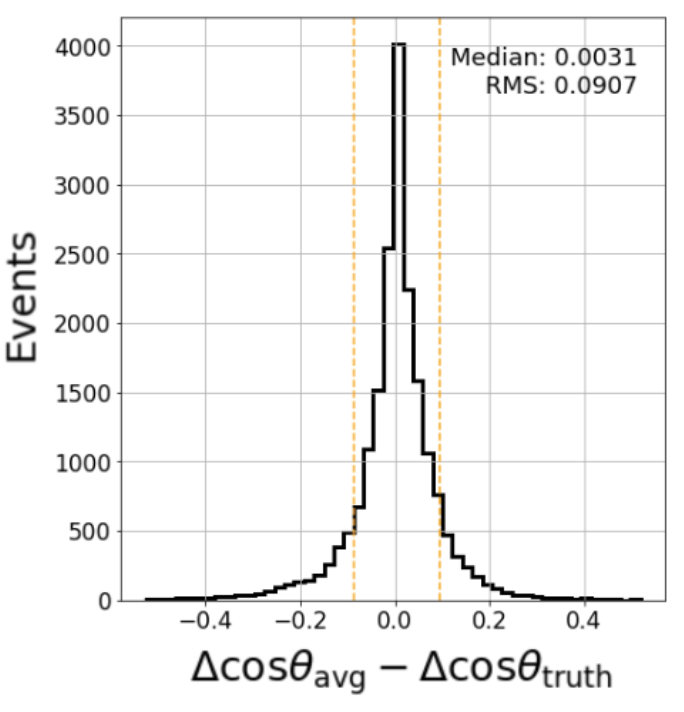}
    \subcaption{}
\end{minipage}
\caption{Reconstruction performance with truth (blue) and average (green) integrated PDFs when a vertex is 15 m from the closest module and the event direction is vertically upward, parallel to a string in an Upgrade-like module spacing; (a) Angular distribution of detected photons relative to the shower axis, (b) Residual timing distributions of detected photons, and (c) Distribution of differences in zenith angle reconstruction resolutions, $\Delta$cos$\theta_\text{avg}$ and $\Delta$cos$\theta_\text{truth}$, when events are reconstructed with the averaged and truth PDF, respectively.}
\label{case_far}
\end{figure}

\indent The features in these PDFs contain information about the incoming directions such as the local geometry around the simulated event in homogeneous ice. Since the scattering process in the ice cannot be solved analytically, these PDFs can only be derived from simulations which means generating new expected distributions for every hypothesis, which soon becomes computationally challenging when reconstructing over a large parameter space. To circumvent the complexity of generating these true PDFs for every hypothesis, we generate representative angular and timing distributions of detected photons using a randomized detector geometry. The procedure is as follows.\\
\indent We randomly position 100 simulated strings, each with 50 modules, within a cylindrical volume of base area 40000 m$^2$ and height 200 m. We set thresholds on the minimum and maximum distances between modules to prevent clustering and large gaps. This approach ensures that the instrumentation density of the randomized geometry approximates that of the Upgrade physics region. The events are then simulated with vertices uniformly distributed within a cylindrical region centered at the geometric center of the instrumentation and maintaining a sufficient distance from the edges to prevent photons from escaping. The fitted spatial and temporal distribution of detected photons across all hit modules and averaged over multiple event vertices are referred to as the vertex-averaged integrated PDFs (Fig.~\ref{PDF_disagreement}; middle panel). The averaging is performed over 10000 events for the 2D histograms (Fig.~\ref{PDF_disagreement}, \ref{case_close}, \ref{case_far}, \ref{PDF_charge}) and over 50000 events for the 3D histograms (Fig. \ref{3D_correlation}) to ensure sufficient MC statistics. The averaging process reduces the photon sampling bias introduced by the local geometry. The integrated PDFs show agreement within the statistical errors  (Fig.~\ref{PDF_disagreement}; right panel) for different incoming direction hypotheses due to the homogeneous ice introducing translational and rotational symmetries.\\
\indent To obtain analytically smooth and continuous functions, variable-bandwidth Kernel Density Estimates (KDE) \cite{VBW_KDE} are fitted to all PDFs in this study. A window moves over all data points and the bandwidth of a Gaussian kernel is calculated depending on the population enclosed within the window. The bandwidth is defined as 1/(population)$^n$ to effectively capture the high-statistics regions while smoothing low-statistics areas. Optimization of the bandwidth and window size parameter involves a Kolmogorov-Smirnov goodness-of-fit test \cite{KS_test} which probes parameter values for window size and bandwidth to minimize differences between histogram bin populations and KDE estimates.

\indent Given that the true integrated PDFs (Fig \ref{PDF_disagreement}; left panel) differ from averaged integrated PDFs (Fig.~\ref{PDF_disagreement}; middle panel), it is important to assess how this modeling error impacts reconstruction resolutions. We therefore evaluate the reconstruction resolutions for two extreme event cases, with cascade energy of 5 GeV, characterized by different PDFs, one with minimal photon scattering and the other dominated by scattering effects in an Upgrade configuration. The true (averaged) integrated PDFs are shown in blue (green) in Fig.~\ref{case_close} and Fig.~\ref{case_far}, with the angular and timing PDFs in the left and middle panels, respectively. We show the 1D projected PDFs without PMT-induced photon direction smearing to highlight the effect of scattering on photon distributions. Note that the subsequent event reconstruction includes all information loss processes, including the photon direction smearing by PMTs. The first case describes a setup where the vertex is 1m away from the nearest module, which is positioned on the Cherenkov ring of the shower. In the second case, the event is directed vertically upwards parallel to a string, with its vertex 15m from the nearest module, resulting in significantly more scattering than in the first scenario. 

\indent We define a resolution bias metric as the difference between the zenith-angle resolutions, $\Delta \text{cos}\theta_\text{avg}$ and $\Delta \text{cos}\theta_\text{truth}$, obtained using the averaged and true integrated PDFs, respectively. Each of the two extreme event scenarios was simulated 10000 times to construct the PDFs while the reconstruction was performed on 20000 independently simulated events. The results in the right panel of Fig.~\ref{case_close} and Fig.~\ref{case_far} show median differences compatible with zero, suggesting that the vertex-averaged PDFs do not introduce an additional bias on average. The RMS spread, which represents the uncertainty introduced by approximating the PDF, is about 0.06 and 0.09 for the two cases respectively. When compared to the overall resolutions (black line) shown in Fig.~\ref{PDF_exclude}, with average hit counts of 50 (20) for the less (more) scattering case, the RMS spread is comparable to the median values of $\sim$0.1 ($\sim$0.13). However, the RMS spread is expected to be smaller in less extreme scenarios where the difference in PDFs is less pronounced. The true integrated angular and timing distributions sum over photon distributions in all modules and are independent of the source-receiver distance. Due to this, they blur the light pattern information in each module. The improvement in resolutions incorporating the distance-dependent angular and timing distributions is discussed later in Section \ref{results} and presented in Fig.~\ref{realvstoy}.

\begin{figure}[!ht]
\centering
\includegraphics[width=0.8\textwidth]{Images/charge_PDF.pdf}
\caption{Contours of a two-dimensional variable-bandwidth KDE fit (solid red lines) to the expected charge for a module at a radius r and angle $\Delta \zeta$ relative to the shower axis for an event, represented as $p_\text{q}$ in Equation \ref{LLH_charge}. The dark bins at the closest distances and large angles from the shower axis indicate zero population and are not indicative of low module charge.}
\label{PDF_charge}
\end{figure}

\subsection{Reconstruction with different observables}
\label{sources_event_info}
\indent Reconstruction is initially performed using the direction (equation \ref{LLH_dir}) and residual timing (equation \ref{LLH_time}) information of photon hits to study them individually. The performance metric here is the resolution, quantified as the absolute difference between reconstructed and true cosine zenith. In NC events, the true direction corresponds to the direction of the hadronic shower axis. In $\nu_e$ CC and $\nu_\tau$ CC events, it lies between the hadronic and electromagnetic shower directions due to momentum transfer between the two shower components \cite{inelasticity}. Error bounds are determined as the $1\mathrm{\sigma}$ resolution uncertainty, derived from the 84th and 16th percentiles of the distribution of residuals for a given photon count. Since the reconstruction resolution is driven by the number of detected photons, we present the resolutions as a function of photon hit counts (instead of cascade energy).\\
\indent To determine the reconstruction resolutions, 30000 purely hadronic events are simulated for each case within a geometry similar to the Upgrade instrumentation, with a uniform module spacing of 20 m horizontally and 3 m vertically. This setup resembles the Upgrade geometry and differs from the previously discussed randomized geometry with randomized string and module positions, which is only used to generate the expected PDFs. The events are confined entirely within the detector with shower energies uniformly distributed between 1-20 GeV with 1500 events per energy bin. The simulations include all the considered information loss processes, as described in Sections \ref{imp_zenith} and \ref{event_sim}. The expected angular and residual timing distribution of hit photons as used in the likelihood PDFs are shown in the middle panel of Fig.~\ref{PDF_disagreement}. The expected charge recorded by a module is modeled by the distribution shown in Fig.~\ref{PDF_charge}. Each event is reconstructed using the different likelihood scenarios presented in \ref{LLH_dir}, \ref{LLH_time}, \ref{LLH_charge} and then by the combined likelihood \ref{LLH_combined_corr}.\\
\indent The zenith angle reconstruction performances using the different observables are shown in Fig.~\ref{info_sources}. Reconstructions that use the direction and timing information of hit photons, shown in dashed-dotted green and dashed brown lines, respectively, are comparable suggesting that both are equally significant sources of directional information. The reconstruction based on the charge information from hit modules, represented by the dotted magenta line, shows relatively poorer resolution. This outcome is expected since the charge represents the photon summary statistics at each module instead of photon-level information, reducing the available information. The expected distribution p$_\text{corr}$ for the combined likelihood is constructed by generating a three-dimensional histogram as a function of the angular distribution of hit PMT vectors and hit modules relative to the shower axis and the residual timing for each hit photon. Three-dimensional KDEs are fitted to the underlying distribution and slices showing the marginalized distributions in each dimension are shown in Fig.~\ref{3D_correlation} for proper visualization. We observe that the improvement in median resolutions, shown by the solid black line, is minimal compared to those obtained when treating photon direction and timing separately. This suggests that leveraging the correlation between observables does not contribute substantial additional information beyond what is captured by the two-dimensional PDFs discussed in Section \ref{photon_dist}.\\
\begin{figure}[!ht]
\centering
\includegraphics[width=0.8
\textwidth]{Images/info_sources.pdf}
\caption{Zenith angle reconstruction resolutions as a function of photon hit counts with unbinned likelihood formulations using (a) only directional information, (b) only residual timing information of individual photon hits, (c) charge information of individual hit modules, and (d) combined information from direction and timing correlation of photon hits. Each data point represents the median resolution for the corresponding bin, with 19 uniformly spaced bins between 5 to 100 photon hits.}
\label{info_sources}
\end{figure}
\indent The black solid line in Fig.~\ref{info_sources} also represents the achievable resolutions with all the considered information loss processes included. They are approximate due to the modeling errors in the averaged integrated PDFs which differ from the true integrated PDFs. Additionally, the hit PDFs represent the integrated photon distributions over all sensors rather than modeled for each sensor, the latter of which might contain additional cascade directional information, as discussed in Section \ref{photon_dist}. The simulation includes all processes that degrade event information, and the likelihood reconstruction is performed with a reasonably good representation of the underlying true integrated PDFs. Given the huge computational cost of generating these PDFs for each module, these resolutions represent the achievable reconstruction performance within computational feasibility. A toy simulation that further improves these resolutions by incorporating the source-receiver distance is discussed in the following section.

\section{Results}
\label{results}
\indent We now investigate the processes that limit the physics information in IceCube Upgrade events and degrade reconstruction resolution. Each event is simulated using the same detector setup discussed in Section \ref{event_sim} and with identical MC information before the photon generation and propagation stage. The subsequent simulation incorporates (or ignores) a specific information loss process one at a time, as detailed in Section \ref{event_sim} while keeping all other processes disabled (or enabled). This simulation step then determines the photon generation, propagation, and detection. Reconstruction is performed using an extended likelihood formulation (Equation \ref{LLH_combined_corr}), which offers the best performance as discussed in Section \ref{sources_event_info} and shown in Fig.~\ref{info_sources}. The expected correlation PDF and module charge, p$_\text{corr}$ and p$_\text{q}$ respectively, are generated for each case with the same information loss processes included as in the respective simulation.

\subsection{Limitations in resolutions from individual information loss processes}

\indent The reconstruction resolutions on the zenith angle as a function of photon hit counts are shown in Fig.~\ref{PDF_include}. The impact of module noise (solid red) on the overall resolution is minimal. This result is expected, given the low dark noise rate in mDOM PMTs resulting in a high signal purity, as discussed in Section \ref{event_sim}. In the absence of scattering effects, the degradation of event information by photon direction smearing by PMTs (dashed green) is also negligible and comparable to module noise. This is because, despite the smearing of the angular distribution of photons caused by the PMTs, the charge distribution still enables an almost perfect reconstruction of the shower axis direction in the absence of a hadronic shower and photon scattering. The degradation in resolution due to the spread of shower particles (dashed-dotted burnt orange) is significantly higher, making it a major factor in limiting event information. The resolution deterioration caused by in-ice photon scattering (dotted blue) is even greater than that from the shower spread highlighting that in-ice scattering is the primary factor restricting the physics information in events.
\begin{figure}[!ht]
\centering
\includegraphics[width=0.8\textwidth]{Images/PDF_include.pdf} 
\caption{Zenith angle reconstruction resolutions as a function of photon hit counts, using the same MC as in Fig.~\ref{info_sources} before photon generation and propagation. Each simulation includes one factor -- (a) module noise, (b) module resolutions, (c) transverse momenta of shower particles, or (d) in-ice photon scattering -- at a time while neglecting other effects. Each data point represents the median resolution for the corresponding bin, with 9 uniformly spaced bins between 5 to 50 photon hits.}
\label{PDF_include}
\end{figure}

\indent The individual impacts of the information loss processes on reconstruction resolutions are analyzed in Fig.~\ref{PDF_include}. Since these processes degrade reconstruction resolution, we evaluate the improvement in resolutions when any one of the resolution-limiting processes is disabled while keeping all others active in the simulation. This approach is the reverse of the earlier study, which analyzed the contribution of each limiting process to the overall resolution. The results in reconstruction resolutions are shown in Fig.~\ref{PDF_exclude}. When the effect of PMT smearing of incident photon directions (dashed green) is removed, the degradation remains negligible indicating that module resolutions have a minimal impact on overall performance. An important implication of these results is that improving on the directional accuracy of the hit photons does not contribute towards further improving our reconstruction. While we benefit from an increase in hit counts due to the larger effective area of the module with more PMTs, this is not relevant to the current study, as resolutions are shown as a function of photon hits rather than energies.\\
\indent Disabling the effect of shower spread (dashed-dotted burnt orange) yields an approximate 30\% improvement in resolutions over the benchmark. The most pronounced improvement, around 50\% over the benchmark, occurs when in-ice photon scattering (dotted blue) is excluded, indicating its dominant influence in limiting resolutions. While the former is inherent to shower formation, the latter could be reduced in a different detection medium with reduced scattering, such as water ~\cite{P-ONE}. The results confirm the conclusions from earlier discussions regarding the contributions of resolution-limiting processes. These systematics in IceCube dominate over the kinematic opening angle (Section \ref{event_sim}; first paragraph) for NC and CC cascade events. 

\begin{figure}[!ht]
\centering
\includegraphics[width=0.8\textwidth]{Images/PDF_exclude.pdf}
\caption{Zenith angle reconstruction resolutions as a function of photon hit counts, with the same MC as in Fig.10 before photon generation and propagation, when a specific information-limiting process is excluded from the simulation. A specific process -- (a) module noise, (b) module resolutions, (c) transverse momenta of shower particles, or (d) in-ice photon scattering -- is excluded at a time while all other factors remain included in the simulation. Each data point represents the median resolution for the corresponding bin, with 14 uniformly spaced bins between 5 to 80 photon hits.}
\label{PDF_exclude}
\end{figure}

\subsection{Achievable resolutions}
\indent The resolutions representing the combined degradation effect on reconstruction from all of the included information loss processes, were derived in Sec.~\ref{photon_dist} and have been shown as a reference line (black dotted) in Fig.~\ref{realvstoy}. To determine the loss in directional resolutions due to PDF modeling errors in this case, a toy simulation is constructed. The angle ($\vec{u},\vec{v}_\text{pmt}$) and residual timings $\Delta$t of hit photons from the PPC simulation are replaced with values sampled from the averaged integrated PDF. The observed values for each photon hit are replaced, ensuring that the integrated PDF and the per-module PDFs replicate the vertex-averaged photon PDFs, shown in the middle panel of Fig. \ref{PDF_disagreement}. Since the observables, $\gamma_{i,s}$ and $t_{\text{obs}_{i,s}}$, are randomly sampled from the averaged PDF, the simulation represents an unphysical scenario where the amount of scattering a photon goes through is distance independent. This means that a module close to an event might register larger timestamps and more scattered photons than a module much further away. However, for this study, this toy simulation simply demonstrates the reconstruction performance when the underlying angular and timing distributions are the same as the expected ones.\\
\indent The results, depicted by the brown solid line in Figure \ref{realvstoy}, indicate that the reconstruction performance improves by approximately 20\% when the underlying photon distributions for each module are perfectly modeled. This outcome aligns with the RMS spread observed when comparing reconstructions using the true and averaged PDFs, as discussed in Section \ref{photon_dist} and shown in the right panel of Figures \ref{case_close} and \ref{case_far}. These insights can serve as a reference for reconstruction algorithms used by the IceCube Collaboration to identify the key processes that limit performance and quantify the potential resolution improvements through a deeper understanding of those processes.
\begin{figure}[!ht]
\centering
\includegraphics[width=0.8\textwidth]{Images/realvstoy.pdf}
\caption{Zenith angle reconstruction resolutions as a function of photon hit counts with the same MC as in Fig.~\ref{info_sources} and Fig.~\ref{PDF_include} before photon generation and propagation, for the simulated events using PPC (dotted black); a more accurate estimation of the achievable resolutions with a toy simulation (solid brown). Each data point represents the median resolution for the corresponding bin. Errors are depicted as shaded regions in the same color as the median, with 19 uniformly spaced bins between 5 to 100 photon hits.}
\label{realvstoy}
\end{figure}
\section{Summary and Conclusions}
\label{summary}

In ice-based neutrino detectors, inhomogeneities and anisotropies within the medium pose challenges (Section \ref{event_sim}) for accurate shower reconstruction since they complicate the modeling of photon propagation through ice. To obtain true likelihood profiles, it is necessary to generate expected angular and timing distributions of detected photons for each direction and vertex position hypothesis, which is computationally impractical for likelihood reconstruction methods.\\
\indent These challenges are circumvented by using homogeneous ice (Section \ref{event_sim}) which ensures that the obtained photon distributions are translationally and rotationally invariant. To bypass the issue of direction-dependent photon distributions in homogeneous ice due to the local detector geometry around the event, a randomized geometry is constructed from which averaged integrated PDFs are generated (Section \ref{reco_method}). These PDFs are independent of incoming directions and are used for all simulated events. An extended unbinned likelihood reconstruction is employed based on the direction and timing information of individual photons along with the per-module charge information, simulated via the Photon Propagation Code (Section \ref{reco_method}) used by the IceCube Collaboration. \\
\indent We have derived the individual contributions from the major processes limiting the physics information, which subsequently impose limitations on the achievable resolutions. We have demonstrated (Fig.~\ref{PDF_include}) that in-ice photon scattering and transverse momentum spread of shower particles (Section \ref{event_sim}) are the dominant factors in limiting physics information. These contributions dominate over limitations arising from the directional resolution of PMTs in modules and module noise, both exhibiting minimal impacts on the reconstruction degradation. Due to negligible topological differences between $\nu_e$ CC, $\nu_\tau$ CC and NC events at these low energies, the contributions from resolution-limiting processes apply to all low-energy cascades observed in IceCube.\\
\indent The achievable reconstruction resolutions in ice are derived using integrated averaged PDFs and a toy simulation further indicates a 20\% improvement in resolutions when angular and timing profiles of detected photons for every module are perfectly modeled instead of using an integrated distribution over all modules. These results are the first step towards estimating the potential improvement in the performance of reconstruction algorithms for GeV neutrino experiments, such as the IceCube Upgrade. Novel machine learning techniques offer improvements over traditional likelihood methods by better accommodating asymmetric detector geometries and thereby boosting reconstruction performance \cite{Jessie_Micalf}\cite{GNN}. However, given the significant computational cost of training these algorithms, which increases with network complexity and benefits from larger training samples, the results from this study provide key insights into optimizing resource allocation to enhance low-energy reconstruction methods.\\
\indent Although the simulations and subsequent reconstruction in this paper have been performed for ice, the analysis is also applicable to water-based Cherenkov detectors like ORCA \cite{KM3Net:2016zxf}. This is because seawater and ice are predominantly H$_2$0 leading to similar neutrino-nucleon cross-sections and subsequent particle propagation and overall shower profiles ~\cite{ORCA_shower}. Therefore, it is expected that the limits on directional resolution by the shower spread will apply equally well in seawater. The proposed detector layout in ORCA has multi-PMT optical modules similar to mDOMs which are deployed on vertical strings. This detector architecture is expected to record photon distributions comparable to those in this study, though they will be shifted due to the different refractive index of water. Due to the lower scattering coefficient of water, photon scattering would have a smaller effect on limiting resolutions in water than in ice.

\pagebreak
\appendix
\section{Supplementary figures and Plots}

\begin{figure}[!ht]
\centering
\includegraphics[width=1.0\textwidth]{Images/3D_correlation.pdf}
\caption{Photon distribution slices from the joint PDF. \textbf{Top row:} Angular distribution slices of hit modules relative to the shower axis, \textbf{Middle row:} Angular distribution slices of hit PMT vectors relative to the shower axis, \textbf{Bottom row:} Slices of residual timing distributions for detected photons, shown relative to the vertex coordinates and time.}
\label{3D_correlation}
\end{figure}

\newpage
\acknowledgments

This work has been supported by the Cluster of Excellence “Precision Physics, Fundamental Interactions, and Structure of Matter” (PRISMA+) funded by the German Research Foundation (DFG) within the German Excellence Strategy.

\printbibliography 
\end{document}